\documentclass{article}
\usepackage{psfig}

\usepackage{scicite}

\usepackage{times}

\newenvironment{sciabstract}{%
\begin{quote} \bf}
{\end{quote}}

\newcounter{lastnote}

\title{Radio Emission from an Ultraluminous X-Ray Source} 

\author {Philip Kaaret$^{1\ast}$, Stephane Corbel$^2$, Andrea H.\
Prestwich$^1$, Andreas Zezas$^1$\\ 
\\
\normalsize{$^{1}$Harvard-Smithsonian Center for Astrophysics, 60
Garden St., Cambridge, MA 02138, USA}
\\ 
\normalsize{$^{2}$Universit\'e Paris 7 \& CEA Saclay (Federation APC), 
CE-Saclay, 91191 Gif sur Yvette Cedex, France}\\
\\
\normalsize{$^\ast$To whom correspondence should be addressed; E-mail: 
pkaaret@cfa.harvard.edu} 
}

\date{}

\newcommand\arcdeg{\mbox{$^\circ$}}% 
\newcommand\arcmin{\mbox{$^\prime$}}% 
\newcommand\arcsec{\mbox{$^{\prime\prime}$}}% 

\begin{document} 

%\baselineskip24pt


\maketitle


\begin{sciabstract}

The physical nature of ultraluminous x-ray sources is uncertain.
Stellar mass black holes with beamed radiation and intermediate mass black
holes with isotropic radiation are two plausible explanations.  We
discovered radio emission from an ultraluminous x-ray source in the
dwarf irregular galaxy NGC 5408. The x-ray, radio and optical fluxes as
well as the x-ray spectral shape are consistent with beamed
relativistic jet emission from an accreting stellar black hole.  If
confirmed, this would suggest that the ultraluminous x-ray sources may
be stellar-mass rather than intermediate mass black holes.  However,
interpretation of the source as a jet-producing intermediate-mass black
hole cannot be ruled out at this time.

\end{sciabstract}

One of the most enigmatic results to emerge from x-ray population
studies of galaxies is the discovery of non-nuclear x-ray sources which
appear to have luminosities factors of 10 to 100's times the Eddington
luminosity for a neutron star -- the so-called ultraluminous x-ray
sources\cite{fabbiano89,colbert99,roberts00}.  The physical nature of
these sources is controversial and possible models include very
luminous supernova remnants\cite{fabian96}, gamma-ray burst
remnants\cite{wang99}, accreting stellar-mass compact objects with
radiation which is beamed mechanically \cite{king01} or
relativistically \cite{kording02}, or accreting intermediate mass black
holes\cite{colbert99} with masses between 100 to several 1000 solar
masses.  The detection of x-ray variability in several sources suggests
that these sources are powered by accretion of material from a
companion star or interstellar gas \cite{makishmia00,kaaret01} and are
not supernova or gamma-ray burst remnants.  If the radiation from these
variable sources is isotropic, then a high inferred luminosity requires
an intermediate-mass black hole.  If the radiation is beamed, then a
stellar-mass black hole could explain the observations. 

%Stellar motions in globular clusters may also suggest the presence of
%intermediate mass black holes \cite{gerssen02}.  

Optical counterpart searches have shown that ultraluminous x-ray
sources preferentially occur near star-forming regions \cite{roberts01}
although they are also found in old globular clusters
\cite{angelini01,wu02}.  In star-forming systems, one unique massive
star counterpart has been identified \cite{liu02} and nebulae have been
detected around several sources \cite{pakull02}.  In some cases,
counterpart searches have shown that the sources are background objects
\cite{foschini02}. Further identification of optical and radio
counterparts is needed to reveal the physical nature of the
ultraluminous x-ray sources.

The dwarf irregular galaxy NGC 5408 contains a well studied
ultraluminous x-ray source, 2E 1400.2-4108
\cite{stewart82,fabian93,motch94,colbert99}.  We observed the galaxy
for a total exposure of 4683~s with the Chandra X-Ray Observatory on 7
May 2002 using the High-Resolution Mirror Assembly and the Advanced
Camera Imaging Spectrometer spectroscopic array operated in an imaging
mode suitable for a high count rate source \cite{chandra_ref}.  Images
in the 0.3--8~keV band showed a point source (we place an upper limit
on the source diameter of $0.9\arcsec$) at a position of $\alpha(\rm
J2000)=$ 14h 03m 19s.63, $\delta(\rm J2000)=$ -41$\arcdeg$ 22$\arcmin$
58$\arcsec$.7 with an error radius of $1\arcsec$ at 90\% confidence and
a flux of $2.7 \times 10^{-12} \rm \, erg \, cm^{-2} \, s^{-1}$ in the
0.3--8~keV band corrected for Galactic absorption along the line of
sight. If the x-ray emission is isotropic, then the source luminosity
would be $1.1 \times 10^{40} \rm \, erg \, s^{-1}$ in the 0.3--8~keV
band at the distance of 4.8~Mpc to NGC 5408 \cite{karachentsev02}.  The
long-term x-ray light curve (Fig.~1) shows variability at a level of
40\%.

A 4.8 GHz radio map of NGC 5408 from the Australia Telescope Compact
Array (ATCA) showed a radio source at the location of the x-ray
source.  We reprocessed the 8 hour ATCA observation of NGC 5408 taken
on 31 March 2000 \cite{stevens02} to produce a natural weighted map
that provides optimum signal to noise ratio for point sources at the
cost of angular resolution.  We detected a 4.8~GHz radio source with a
flux of $0.26 \pm 0.04 \rm \, mJy$ (1$\sigma$ error) at a  position of
$\alpha(\rm J2000)=$ 14h 03m 19s.606 and $\delta(\rm J2000)=$
-41$\arcdeg$ 22$\arcmin$ 59$\arcsec$.572 (J2000) with an error radius
of $0.1\arcsec$ at 90\% confidence \cite{reynolds95}, consistent with
the Chandra position within the astrometric uncertainties.  No source
is present at the same location in 8.64~GHz data obtained
simultaneously with the 4.8~Ghz data and we place an upper limit
(3$\sigma$) on the flux of 0.12~mJy based on the rms noise near that
location.  From the 4.8~GHz detection and the 8.64~GHz upper limit, we
place a lower bound on the radio spectral index $\alpha_{r} > 1.0$,
defined so $S_{\nu} \propto \nu^{-\alpha_{r}}$.  From radio source
counts\cite{partridge86} at 4.9~GHz, the probability of chance
occurrence of a 0.18~mJy or brighter background radio source within the
Chandra error circle is less than $1.0 \times 10^{-5}$.  To estimate
the probability of chance coincidence with a radio source internal to
the galaxy, we compare with NGC 1569 which is a dwarf irregular similar
to NGC 5408 and has two sources which would have a 4.8~GHz flux of
0.18~mJy or larger if placed at the distance to NGC 5408
\cite{greve02}.  The probability of chance coincidence within
$1\arcsec$ of the Chandra source given 2 radio sources within the
optical extent of NGC 5408 is less than 0.002.  Further, all of the
radio sources in NGC 1569 have $\alpha_{r} < 0.6$, which is
inconsistent with the steep spectrum of our radio source.

Several optical sources in a Hubble Space Telescope (HST) image of NGC
5408 (Fig.~2) lie within the ATCA/HST error circle; note that the
relative error circle is dominated by the HST astrometric uncertainty. 
The brightest object in the radio or x-ray error circles has a visual
magnitude of 22.1.  The colors of the objects detected in both bands
are similar to those of the overall stellar population of the galaxy
\cite{karachentsev02}.  None of the objects show evidence of extended
emission.  

Given the radio counterpart and the brightest optical counterpart in
the Chandra error circle (which we take as an upper limit to the
optical flux), we find that the x-ray to optical flux ratio
\cite{maccacaro82} is $f_x/f_v > 380$.  We can also calculate two-point
spectral indices\cite{tananbaum79,stocke91} between the radio, optical,
and x-ray bands using the flux density at 4.8~GHz, the V band, and
1~keV.  We find that the radio to x-ray index $\alpha_{rx} = 0.16$, the
optical to x-ray index $\alpha_{ox} < 0.18$, and the radio to optical
index $\alpha_{ro} > 0.15$.  The ratio $f_x/f_v$ of the object is much
larger than the ratios found for stars, normal galaxies, clusters of
galaxies, and AGN\cite{maccacaro82,stocke91}.  We rule out the
possibility that the object is a member of any of these classes of
source.  Identification as a blazar is ruled out because the $f_x/f_v$
exceeds that of known blazars \cite{stocke91,landt01}, the radio
spectrum is inconsistent with those of blazars which have $\alpha_{r} <
0.8$ \cite{landt01}, and the x-ray to radio flux ratio is $9\times
10^{-9}$ which is an order of magnitude larger than the most extreme
known blazar \cite{landt01}.

Some young supernova remnants have high x-ray luminosities.  However, 
supernova remnants produce copious optical line emission, which is not
detected here.  Taking the brightest source in the Chandra error
circle, the upper limit on the H$\alpha$ emission is $7 \times 10^{-15}
\rm \, erg \, cm^{-2} \, s^{-1}$ if all of the flux in the HST F606W
filter is H$\alpha$ emission.  A similarly obtained upper limit on the
[O{\sc iii}] emission is $1.6 \times 10^{-14} \rm \, erg \, cm^{-2} \,
s^{-1}$.  These upper limits are below the line fluxes measured for
highly x-ray luminous supernova remnants \cite{chu01,blair83}.  Optical
line emission could be hidden by a high obscuring column
\cite{roberts02}, but such obscuration is inconsistent with the
relatively low absorbing column density found from the x-ray spectrum
(Fig.~3).  Also, the observed x-ray variability would be unexpected for
a supernova remnant.  The line fluxes and the $f_{x}/f_{v}$ ratio are
also inconsistent with identification of the source as a young
supernova.

X-ray binaries and rotation-powered neutron stars sometimes have high
$f_x/f_v$ ratios similar to what we find for the source in NGC 5408.  
The x-ray variability indicates the source is not a rotation-powered
neutron star.  If the source was located in the Milky Way, the
luminosity would be of order $10^{33} - 10^{34} \rm \, erg \, s^{-1}$. 
Radio emissions when detected from Galactic x-ray binaries in such low
luminosity states \cite{fender01,corbel01} have a flat or inverted
spectra which is inconsistent with the steep spectrum of the source and
suggests that the source is not a low-luminosity x-ray binary in the
Milky Way.  The steep radio spectrum is consistent with optically thin
synchrotron emission as produced by x-ray binaries in high luminosity
states or during state transitions \cite{mirabel99,fender99,corbel01}
and is consistent with the source being an x-ray binary in NGC 5408.
The source x-ray variability may be less than seen from many x-ray
binaries, but is consistent with that of Cyg X-1.

Radio emissions from x-ray binaries are associated with relativistic
jets.  The radio emission we detected from the ultraluminous x-ray
source in NGC 5408 is probably from a relativistic jet.  It could be
from an accreting stellar-mass black hole producing relativistic jets
(a microquasar) beamed toward us \cite{mirabel99}.  The intensity of
isotropic rest-frame emission from a continuous jet with a power law
spectrum with radio spectral index $\alpha_r$ is amplified by a factor
$\delta^{2+\alpha_r}$ where $\delta = \gamma^{-1}(1-\beta \cos
\theta)^{-1}$ is the Doppler factor (for the approaching jet) where
$\beta$ is the jet speed as a fraction of the speed of light and
$\gamma = (1-\beta^2)^{-0.5}$.  The Galactic microquasar GRS 1915+105
has $\delta = 0.34$ and a distance near $11 \rm \, kpc$
\cite{fender99}.  GRS 1915+105 is highly variable.  Radio fluxes of
about $300 \rm \, mJy$ at 4.8~GHz are observed in bright radio flares
when superluminal jets are resolved, whereas the more typical radio
fluxes are about $10 \rm \, mJy$ while the source is x-ray active but
not undergoing a bright radio flare.  If the ATCA observation happened
to catch the NGC 5408 source during such a bright flare, then a Doppler
factor as low as $\delta = 2.1$ would suffice to produce the observed
radio flux (with $\alpha_r = 1.0$).  However, the probability of
catching such an event with a single radio observation is low unless
the NGC 5408 source is an extremely active jet source.  Scaling from
the more typical radio flux of GRS 1915+105 of about $10 \rm \, mJy$,
we find that a Doppler factor $\delta > 5.8$ would be needed to produce
the observed flux.  This is compatible with the constraints on the
ejecta velocities for GRS 1915+105.  If $\gamma = 5$ as inferred for
GRS 1915+105, then tilts of the jet axis relative to the line of sight
of up to $10\arcdeg$ could produce $\delta > 5.8$.  Roughly 1 out of 70
of extragalactic microquasars should be aligned like this.  A steeper
radio spectral index or higher intrinsic luminosity would reduce the
required $\delta$.

The x-ray emission could also be relativistically beamed jet emission
\cite{kording02}. The fact that $\alpha_r > \alpha_{rx}$ rules out the
possibility that a single component spectrum, like synchrotron
emission, produces both the radio and x-ray emission. 
Inverse-Comptonization of photons from a massive companion star by a
jet may power the ultraluminous x-ray sources\cite{geo02}.  The x-ray
spectrum predicted by this process has the form of a broken power law
with the break near 1~keV and a cutoff at energies above the x-ray
band.  The Chandra x-ray spectrum is consistent with this form (Fig.~3)
with a break energy of $0.65 \pm 0.06 \rm \, keV$ and a spectral index
at high energies of $\alpha_x = 2.1 \pm 0.2$.  The beaming pattern for
inverse-Compton emission with a photon source external to the jet,
e.g.\ from the companion star, varies as $\delta^{3+2\alpha_x}$ for a
continuous jet, so very large amplifications are possible, e.g.\
$4\times 10^5$ for $\delta = 6$.  Hence, the jet x-ray luminosity may
be as low as $10^{34} \rm \, erg \, s^{-1}$.  The break energy we
measure is similar to what has been modeled \cite{geo02}.  However, our
measured x-ray spectral parameters would imply an unusually steep power
law index for the injected relativistic electrons of $p = 4.2 \pm 0.4$.

If the source is an accreting black hole and the x-ray emission is
isotropic, then the luminosity implies a minimum mass of $80
M_{\odot}$. The spectral index of the hard spectral component, if the
spectrum is fitted with the sum of a power law plus thermal accretion
disk emission, is $\alpha_x = 1.75 \pm 0.34$ which is similar to that
found for black hole candidate x-ray binaries in their high flux states
\cite{tanaka96}.  The temperature of the soft thermal component is $kT
= 0.11 \pm 0.03 \rm \, keV$ which is lower than found for stellar mass
black holes in the Milky Way \cite{ebisawa94}.  Such a low temperature
is also unlikely to arise from mechanically beamed radiation from a
stellar-mass black hole \cite{king01}.  Some other ultraluminous
sources have high temperatures, which is inconsistent with the presence
of intermediate mass black holes unless they are very rapidly rotating
\cite{makishmia00}.  The low temperature found for the NGC 5408 source
is consistent with that expected for a non-rotating black hole with a
mass above $100 M_{\odot}$ \cite{makishmia00,colbert99}.  The radio
spectral index and the radio to x-ray flux ratio are similar to the
(highly variable) ones measured for microquasars
\cite{mirabel99,fender99} suggesting that the radio emission could
arise from jets similar to those produced by microquasars.  The source
lies about $12\arcsec$ from the star formation regions of NGC 5408 that
contain super star clusters, giving a projected displacement of
280~pc.  An x-ray binary with a speed of 10~km/s, from either
interactions in a cluster or a supernova kick, could traverse this
distance in 30~My.  Hence, an intermediate mass black hole could have
formed in one of the observed super star clusters \cite{ebisuzaki01}
and moved to the present location within the life time of a massive
companion star.

The ultraluminous x-ray sources have been interpreted as evidence for
intermediate mass black holes \cite{colbert99}.  The results presented
here show that the radio, optical, and x-ray properties of the
ultraluminous x-ray source in NGC 5408 are consistent with beamed
emission of a relativistic jet from a stellar-mass black hole.  This
suggests that, at least some of, the ultraluminous x-ray sources may be
beamed emission from stellar-mass black holes \cite{kording02}.

%--------------

\begin{center} \begin{figure}
\centerline{\psfig{file=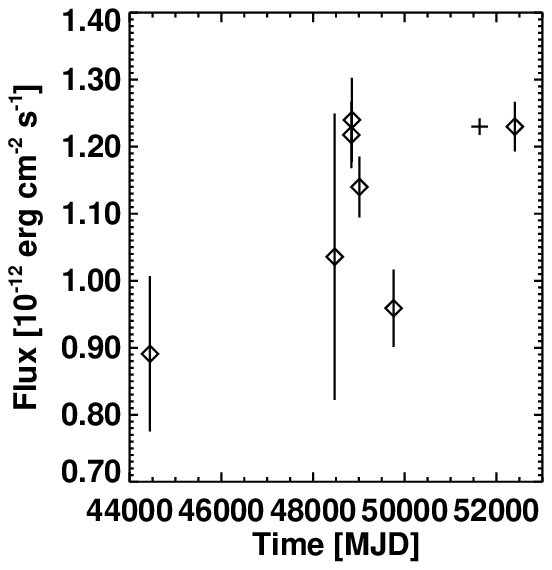,height=3in}} \caption{ Long term x-ray
light curve for the x-ray source in NGC 5408.  Data shown as diamonds
are, in chronological order, from the Einstein IPC\cite{harris93}, the
ROSAT all-sky survey, the ROSAT HRI\cite{motch94}, two observations
with the ROSAT PSPC\cite{stewart82}, the ASCA SIS\cite{colbert99}, and
Chandra.  The cross marks the time of the ATCA observation (the flux is
arbitrary). The absorbed flux in the 0.5--2~keV band, calculated from
each instrumental count rate using a spectral model derived from
fitting the Chandra data, is plotted.  The source appears variable with
a ratio between the maximum and minimum fluxes of 1.4. } \label{lc}
\end{figure} \end{center}

\begin{center} \begin{figure}
\centerline{\psfig{file=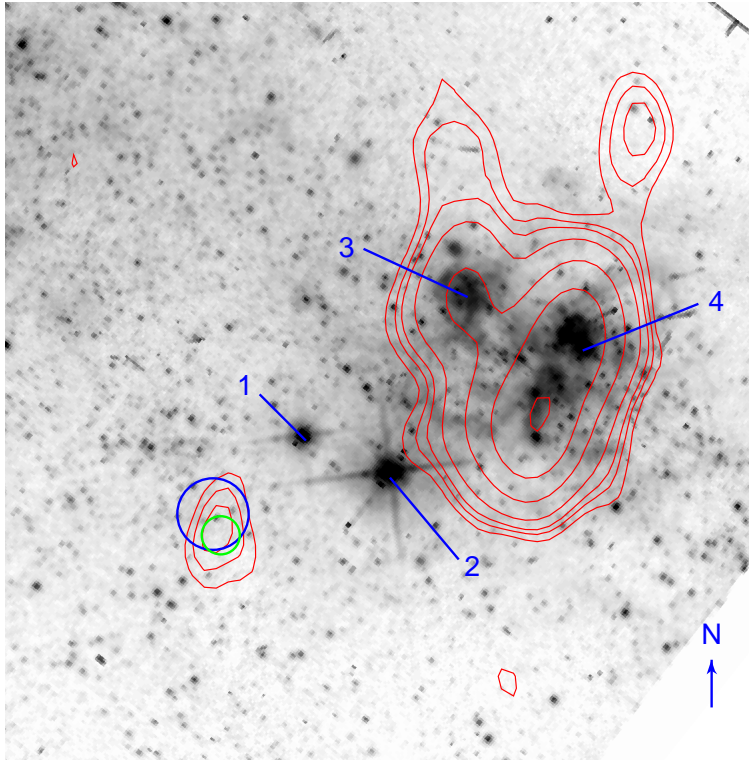,height=4in}} \caption{An
optical image of NGC 5408 showing the region containing the
ultraluminous x-ray source with the ATCA 4.8~GHz radio contours (in red
at flux densities of: 0.12, 0.16, 0.2, 0.3, 0.6, 1.2, and 2.4
mJy/beam), error circles for the Chandra (blue) and ATCA (green)
sources, and the main star formation regions (marked in blue as 3 and
4, note the associated radio emission\cite{stevens02}; 1 and 2 are
foreground stars\cite{motch94}).  The arrow points North and has a
length of $2\arcsec$.  To allow for the various astrometric
uncertainties, we used a $1.5\arcsec$ radius circle around the Chandra
position and a $0.8\arcsec$ radius circle around the ATCA position to
identify possible optical counterparts.  Two HST WPFC2 exposures
\cite{karachentsev02}, both 600~s and obtained on 4 July 2000, were
analyzed using the HSTphot\cite{dolphin00} stellar photometry package
to obtain simultaneous photometry in the two filters (F814W and F606W)
and corrected for reddening using an extinction $\rm E(B-V) = 0.068$
based on dust maps\cite{schlegel98} and an $\rm R_V = 3.1$ extinction
curve.  The image shown is with the F606W filter which includes the
lines H$\alpha$ and [O{\sc III}].  We corrected the absolute astrometry
of the HST images using stars from the USNO A2.0 catalog and {\it
imwcs} tool from the Smithsonian Astrophysical Observatory Telescope
Data Center.  There are no USNO A2.0 stars  on the WF3 chips where the
x-ray source is located, and we estimate a total astrometric error for
positions on the WF3 chip of $0.5\arcsec$. } \label{hst_image}
\end{figure} \end{center}

\begin{center} \begin{figure}
\centerline{\psfig{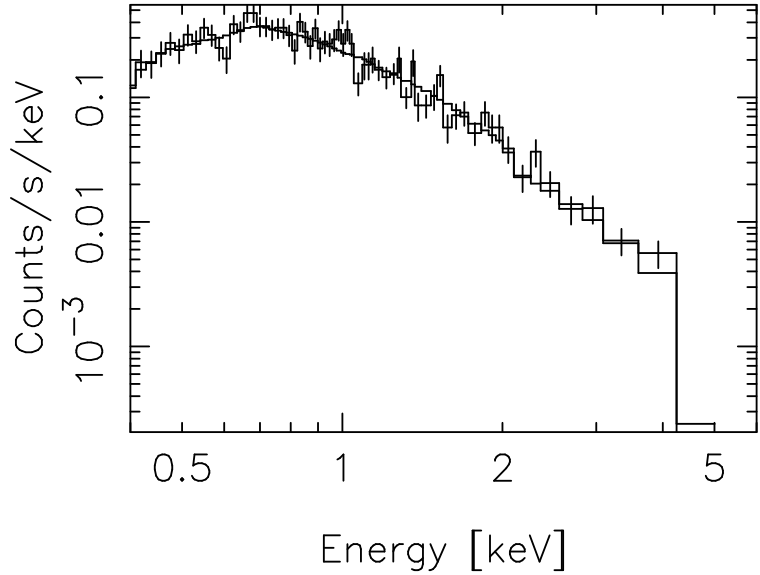}} \caption{X-ray
spectrum for the ultraluminous x-ray source in NGC 5408 in the 0.4--8
keV band.  The x-ray spectrum is inconsistent with simple
single-component models.  A broken powerlaw model with an absorbing
column density equal to the Galactic line-of-sight value of $5.6 \times
10^{20} \rm \, cm^{-2}$ gave a good fit with $\chi^2/\rm DoF =
62.5/61$, a low energy powerlaw photon index between -0.8 and +1.7, a
break energy of $0.65 \pm 0.06 \rm \, keV$, and a high energy powerlaw
photon index of $3.1 \pm 0.2$.  The spectrum is also adequately fitted
($\chi^2/{\rm DoF} = 62.6/60$) with a model consisting of thermal
emission from an accretion disk at a temperature $kT = 0.11 \pm 0.03
\rm \, keV$  with an added powerlaw component with photon index $\Gamma
= 2.75 \pm 0.34$ and an absorption column density $N_{\rm H} = (3.5 \pm
1.9) \times 10^{21} \rm \, cm^{-2}$, significantly above the Galactic
column.  Here and in the text, all error quoted on spectral parameters
are 90\% confidence for a single parameter.  The spectral modeling was
done with the ISIS software (http://space.mit.edu/ASC/ISIS) to correct
for the moderate (14\%) pileup due to the high counting rate.}
\label{chandra_spec} \end{figure} \end{center}

\end{document}